\documentclass[twocolumn,pre,showpacs]{revtex4}
\usepackage[english]{babel}

\usepackage[active]{srcltx}

\usepackage{graphics,graphicx,dcolumn,bm,fleqn,epic,eepic,float,dsfont}
\usepackage{amssymb,amsmath,multirow,rotate,color}
\usepackage{subfigure}

\newcommand{\bA     }{\mbox{\boldmath$A$}}
\newcommand{\bM     }{\mbox{\boldmath$M$}}

\newcommand{\bI     }{\mbox{\boldmath$I$}}

\newcommand{\bphi     }{\mbox{\boldmath$\phi$}}
\newcommand{\bpsi     }{\mbox{\boldmath$\psi$}}

\usepackage{color}

\begin{document}
\title{Index statistical properties of sparse random graphs}
\author{F. L. \surname{Metz}$^{1,2}$ and Daniel A. \surname{Stariolo}$^{2}$}
\affiliation{$^1$ Departamento de F\'isica, Universidade Federal de Santa Maria, 97105-900 Santa Maria, Brazil \\ 
$^2$ Departamento de  F\'isica, Universidade Federal do Rio Grande do Sul, 91501-970 Porto Alegre, Brazil}

\begin{abstract}
Using the replica method, we develop an analytical approach to compute the
characteristic function for the probability $\mathcal{P}_N(K,\lambda)$ that a large $N \times N$ adjacency 
matrix of sparse random graphs has $K$ eigenvalues below a threshold $\lambda$.
The method allows to determine, in principle, all 
moments of $\mathcal{P}_N(K,\lambda)$, from which the typical
sample to sample fluctuations can be fully characterized.
For random graph models with localized eigenvectors,
we show that the index variance 
scales linearly with $N \gg 1$ for $|\lambda| > 0$, with a model-dependent prefactor
that can be exactly calculated. Explicit results  are 
discussed for Erd\"os-R\'enyi and regular
random graphs, both exhibiting a prefactor with a non-monotonic behavior
as a function of $\lambda$.
These results contrast with rotationally invariant random matrices, where the index variance 
scales only as $\ln N$, with an universal prefactor that is independent of $\lambda$.
 Numerical diagonalization results confirm the exactness of
our approach and, in addition, strongly support the Gaussian nature
of the index fluctuations.
\end{abstract}


\maketitle


\section{Introduction} 

Since the pioneering work of Wigner in the statistics of nuclear 
energy levels \cite{Wigner}, random matrix theory has established itself as a research field
on its own, with many important applications in physics
and beyond \cite{mehta}. Valuable information on the behavior
of different systems may be extracted from the eigenvalue statistics of related random matrix 
models. In this respect, meaningful statistical observables are the eigenvalue
distribution, the distribution of extreme eigenvalues and the nearest-level spacing
distribution, to name just a few \cite{mehta}.

Another prominent observable is
the index $\mathcal{K}_N(\lambda)$ of a $N \times N$ random matrix, defined
here as the total number of eigenvalues below a threshold $\lambda$. The random variable $\mathcal{K}_N(\lambda)$
is of fundamental importance in the characterization of disordered systems described by a potential energy surface $\mathcal{H}(x_1,\dots,x_N)$
in the $N$-dimensional configurational space \cite{wales}. 
The eigenvalues of the symmetric Hessian matrix $\bM$, formed by the second derivatives
$M_{ij} = \partial^{2} \mathcal{H}/\partial x_i \partial x_j$, encode all
information regarding the stability properties. The number of positive (negative) eigenvalues
counts the number of stable (unstable) directions around a certain configuration, while
the magnitude of an eigenvalue quantifies the surface curvature along the corresponding direction.
In particular, the minima (maxima) 
of the potential energy are stationary points in which all Hessian eigenvalues are positive (negative).
The index is a valuable tool to probe the energy landscape of
systems as diverse as liquids \cite{Angelani,Broderix}, spin-glasses \cite{Kurchan,Cavagna,Daniel}, synchronization
models \cite{Dhagash} and biomolecules \cite{wales}.

The simplest model for the Hessian of a disordered system consists in neglecting its dependency 
with respect to the configurations and assuming that the elements $M_{ij}$ are independently drawn
from a Gaussian distribution. In this case, the Hessian belongs to the GOE ensemble
of random matrices \cite{mehta}
and the index statistics has been studied originally in reference \cite{Cavagna}, using a fermionic 
version of the replica method. The authors have obtained the large-$N$ behavior of the index 
distribution $\mathcal{P}_N(K,\lambda)$ 
\begin{equation}
\mathcal{P}_N(K,\lambda) \sim \exp{\Bigg{\{} - \frac{\pi^{2}}{ 2 \ln N} \left[ K - N m(\lambda)    \right]^{2} \Bigg{\}}   } ,
\label{kkbba}
\end{equation}
where $m(\lambda) = \int_{0}^{\lambda} d \lambda^{\prime} \rho(\lambda^{\prime} )$ follows
from the Wigner semi-circle law \cite{mehta} for the eigenvalue distribution $\rho(\lambda )$. 
Equation (\ref{kkbba})
implies that, for $N \gg 1$, the index variance scales logarithmically with $N$ and the {\it typical}
fluctuations on a scale of width $O(\sqrt{\ln N})$ around the average index have a Gaussian form.

Recently, a significant amount of work has been devoted to study the index distribution
of rotationally invariant ensembles, including Gaussian \cite{Majumdar1,Majumdar2}, Wishart \cite{Vivo} and Cauchy
random matrices \cite{Majumdar3}. These models share the property that the joint probability distribution
of eigenvalues is analytically known, which allows to employ the Coulomb gas technique, pioneered
by Dyson \cite{dyson}, to compute not only the typical index distribution, but also 
its large deviation regime, which characterizes {\it atypical} large fluctuations  \cite{ Majumdar1,Majumdar2,Vivo,Majumdar3}. 
For all these ensembles, eq. (\ref{kkbba}) is
recovered in the regime of small fluctuations, 
with a variance that grows as $\sigma^2 \ln N$ for large $N$. The prefactor $\sigma^2$ 
is given by $\sigma^2 = 1/\pi^{2}$ for both Gaussian \cite{Cavagna,Majumdar1,Majumdar2}
and Wishart \cite{Vivo} random matrices, independently of $\lambda$, while $\sigma^2 = 2/\pi^{2}$ for Cauchy
random matrices \cite{Majumdar3}. 
This logarithmic behavior of the variance apparently reflects 
the repulsion between neighboring levels \cite{stock}, which imposes a 
constraint on the total number of eigenvalues that fit in a finite
region of the spectrum.

Despite the success of the Coulomb gas approach, the analytical form
of the joint probability distribution of eigenvalues is not known for various
interesting random matrix models.
Perhaps the most representative example in this sense is 
the adjacency matrix of sparse random graphs \cite{bollobas,wormald}, in which the average total number of nonzero 
entries scales only linearly with $N$. Although the eigenvalue distribution
of random graphs has been computed using different techniques
\cite{TimTese}, the statistical properties
of the index have not been addressed so far.
Several random graph models typically contain localized eigenvectors
at  finite sectors of the spectrum \cite{fyodorov,Monasson,Metz2010,Slanina2012}, usually corresponding 
to extreme eigenvalues, where
the nearest-level spacing distribution follows a Poisson law \cite{Slanina2012,Mendez}.
In these regions, neighboring eigenvalues are
free to be arbitrarily close to each other, which should heavily influence the index 
fluctuations. 
Models in which the state variables
are placed on the nodes of random graphs have found an enormous number
of applications, including 
spin-glasses, satisfiability problems, error-correcting codes and complex networks (see \cite{Mezardbook,Barrat}
and references therein), and alternative tools to study their index fluctuations would
be  more than welcome.

In this paper 
we derive an analytical expression
for the characteristic function of the index distribution describing 
the adjacency matrix of a broad class of random graphs, defined in terms of an arbitrary
degree distribution. 
In principle, such analytical result
allows to calculate the leading contribution in the large-$N$ limit of all moments of $\mathcal{P}_N(K,\lambda)$, yet we concentrate here on the first and second
moments. Specifically, we show that the index variance of random
graphs scales generally as $\sigma^{2}(\lambda) N$, with a prefactor $\sigma^{2}(\lambda)$
that depends on the threshold $\lambda$ and on the particular structure of the random graph model
at hand. 
For random regular graphs with uniform edges, in which all eigenvectors are 
delocalized \cite{jakobson, Smilansky,Geisinger}, we show that $\sigma^{2}(\lambda) = 0$ for any $\lambda$.
On the other hand, for random graph models with localized eigenvectors \cite{K08,Metz2010,Slanina2012,Mendez,semerjian},
the prefactor $\sigma^{2}(\lambda)$  exhibits a maximum for a
certain $\lambda$, while it vanishes for $|\lambda| \rightarrow 0$. 
These results indicate that the linear scaling of the variance is a consequence of
the uncorrelated  nature of the eigenvalues in the
localized regions of the spectrum.
Since $\sigma^{2}(0) = 0$ for random graphs with an arbitrary degree distribution, 
the linear scaling breaks down for $\lambda = 0$ and the logarithmic scaling reemerges
as the large-$N$ leading contribution for the index variance, which is supported by numerical diagonalization
results. The model-dependent character of $\sigma^{2}(\lambda)$ contrasts with the highly universal prefactor  
found in rotationally invariant ensembles, though
the typical index fluctuations of random graphs remain Gaussian distributed, as supported
by numerical diagonalization results.

In the next section, we lay the ground for the replica computation of the characteristic function.
The random graph model is introduced  in section \ref{sec2}, the replica approach
is developed in section \ref{sec3} and the final analytical result for the characteristic function is
presented in section \ref{sec4}. We discuss explicit results for the average and the variance
of the index in section \ref{sec5} and, in the final section, some final remarks
are presented. 


\section{The general setting} \label{sec1}

In this section we show how to recast the problem of computing the
index distribution of a random matrix in terms of a calculation reminiscent
from the statistical mechanics of disordered systems.
Let us consider a $N \times N$ real symmetric matrix $\bA$ 
with eigenvalues $\lambda_1, \dots,\lambda_N$. The density of eigenvalues between $\lambda^{\prime}$
and $\lambda^{\prime}+ d \lambda^{\prime}$ reads
\begin{equation}
\rho_N(\lambda^{\prime}) = \sum_{\alpha=1}^{N} \delta(\lambda^{\prime} - \lambda_\alpha) \,.
\end{equation}
The index is defined here as the total number of eigenvalues smaller
than a threshold $\lambda$
\begin{equation}
\mathcal{K}_N(\lambda) = \int_{-\infty}^{\lambda} d \lambda^{\prime} \rho_N(\lambda^{\prime}) = 
\sum_{\alpha=1}^{N} \Theta(\lambda-\lambda_{\alpha}),
\label{i1}
\end{equation}
where $\Theta(\dots)$ is the Heavside step function.
The object $\mathcal{K}_N(\lambda)$ is also regarded as the integrated density of states or the cumulative
distribution function. 
At this point we introduce the generating function
\begin{equation}
\mathcal{Z}_N(z) 
= \left( \frac{-i}{2 \pi} \right)^{\frac{N}{2}}
\int d \bphi \exp{\left[\frac{i}{2} \bphi^T . \left(\bA - \bI z   \right) \bphi   \right]},
\label{ewk}
 \end{equation}
with $\bphi=(\phi_1,\dots,\phi_N)$ and $z=\lambda - i \epsilon$, where $\epsilon > 0$ is a regularizer that
ensures the convergence of the above Gaussian integral
and $\bI$ denotes the identity matrix. The vector components $\phi_1,\dots,\phi_N$ are
real-valued.
By using an identity that relates the Heavside function with 
the complex logarithm, eq. (\ref{i1}) can be written in terms of $\mathcal{Z}_N(z)$ as follows
\begin{equation}
\mathcal{K}_N(\lambda) = \frac{1}{\pi i } \lim_{\epsilon \rightarrow 0^{+}} \left[  \ln \mathcal{Z}_N(z^{*}) - \ln \mathcal{Z}_N(z)  \right]\,.
\label{gh}
\end{equation}
Equation (\ref{gh}) holds for a single matrix $\bA$ with an arbitrary dimension $N$. 

An ensemble of random matrices is defined by a large set of instances 
of $\bA$ drawn independently from a distribution $p(\bA)$.
In this paper, we are interested in computing 
the averaged index distribution
\begin{equation}
\mathcal{P}_N (K,\lambda) = \left\langle  \delta \left[ K - \mathcal{K}_N(\lambda)  \right] \right\rangle,
\label{ghs}
\end{equation}
where $\langle \dots \rangle$ denotes the ensemble average with $p(\bA)$.
Using an integral representation
of the Dirac delta and substituting eq. (\ref{gh}) in eq. (\ref{ghs}), we obtain
\begin{equation}
\mathcal{P}_N (K,\lambda) = \int \frac{d \mu}{2 \pi} e^{- i \mu K} \mathcal{G}_N (\mu,\lambda)  \,,
\label{fg1}
\end{equation}
where the characteristic function
\begin{equation}
\mathcal{G}_N (\mu,\lambda) = \lim_{\epsilon \rightarrow 0^{+}} 
\left\langle  \left[ \mathcal{Z}_N(z)    \right]^{-\frac{\mu}{\pi}}  \left[ \mathcal{Z}_N(z^{*})    \right]^{\frac{\mu}{\pi}}    \right\rangle  
\label{fg2}
\end{equation}
contains the whole information
about the statistical properties of the index.
The moments of the index distribution are determined from
\begin{equation}
\langle K^{n} \rangle = (- i)^{n} \frac{\partial^{n} \mathcal{G}_N (\mu,\lambda)}{\partial \mu^{n}} \Bigg{|}_{\mu = 0}, \quad n \in \mathbb{N}.
\label{mom}
\end{equation}
The aim here is to compute the leading contribution to $\mathcal{G}_N (\mu,\lambda)$ for $N \rightarrow \infty$. 
According to eq. (\ref{fg2}), $\mathcal{G}_N (\mu,\lambda)$ is calculated from the 
ensemble average of a function that contains
real powers of the generating function, which is an unfeasible computation. In order
to proceed further, we invoke the main strategy of the replica method and rewrite eq. (\ref{fg2})
as follows 
\begin{equation}
\mathcal{G}_N (\mu,\lambda) = \lim_{\epsilon \rightarrow 0^{+}}  \lim_{n_{\pm} \rightarrow \pm \frac{\mu}{\pi} }    
\left\langle \left[ \mathcal{Z}_N(z)    \right]^{n_{-}}  \left[ \mathcal{Z}_N(z^{*})    \right]^{n_{+}} \right\rangle  \,.
\label{fra}
\end{equation}
The idea is to treat initially   $n_{-}$ and $n_{+}$ as integers, which allows to
compute the ensemble average. Once this average is calculated and the limit 
$N \rightarrow \infty$ is taken, we make an analytical continuation of $n_{\pm}$
to the real values $\pm \frac{\mu}{\pi}$.


\section{Random graphs with an arbitrary degree distribution}  \label{sec2}

We study the index distribution of $N \times N$ symmetric adjacency
matrices with the following entries 
\begin{equation}
A_{ij} = c_{ij} J_{ij}, 
\end{equation}
where $c_{ij} = c_{ji}$ and $J_{ij} = J_{ji}$.
The variables $c_{ij} \in \{ 0,1 \}$ encode the topology of the underlying random graph: we
set  $c_{ij} = 1$ if there is an edge between nodes $i$ and $j$, and zero
otherwise. 
The real variable $J_{ij}$ denotes the
weight or the strength of the undirected coupling between the adjacent nodes $i$ and $j$.

Both types of random variables are drawn independently
from probability distributions. At this stage, there is no need to specify the distribution $P(J)$ of 
the entries $J_{ij}$ and the model
definitions are kept as general as possible. 
However, we do need to specify the distribution of $\{ c_{ij} \}$, which is given by \cite{Zechina2002}
\begin{eqnarray}
p(\{ c_{i < j} \}) &=& \frac{1}{C_N} \prod_{i < j} \left[ \frac{c}{N} \delta_{c_{ij},1} + \left(1 -   \frac{c}{N}\right) 
\delta_{c_{ij},0} \right] \nonumber  \\
&\times&
\left[\prod_{i=1}^{N} \delta_{k_i,\sum_{j=1}^{N}c_{ij}}  \right], \quad c_{ii} = 0, \label{prs}
\end{eqnarray}
where the product $\prod_{i < j}$ runs over all distinct pairs of nodes and
$C_N$ is the normalization factor.

In this model, the topology of the corresponding graph is solely
determined by the degree $k_i(\{ c_{i < j} \}) = \sum_{j=1}^{N} c_{ij}$ of each node $i$, defined as the total number 
of edges attached to $i$. According to eq. (\ref{prs}), any two nodes are connected
with probability $c/N$, in which $c$ is the average degree, while 
the term involving the Kronecker delta ensures
that the number of edges attached to a certain node $i$ is constrained
to an integer $k_i$.
For $N \rightarrow \infty$, averaged quantities with
respect to $p(\{ c_{i < j} \})$ should depend only upon the degree distribution 
\begin{eqnarray}
p_k = \lim_{N \rightarrow \infty} \frac{1}{N} \sum_{i=1}^{N}   \delta_{k,k_i} \,.
\end{eqnarray}
Equation (\ref{prs}) comprises a large class of random
graph models with distinct degree distributions, provided they fulfill 
$c = \sum_{k=0}^{\infty} p_k k$.
Although the ensemble average in the replica approach is performed 
with the distribution of eq. (\ref{prs}) and the final expression for $\mathcal{G}_N (\mu,\lambda)$
is presented in its full generality, we discuss in section \ref{sec5}
explicit results for regular and 
Erd\"os-R\'enyi (ER) random graphs, where the degree distributions are given, respectively, by $p_k = \delta_{k,c}$ \cite{wormald}
and $p_k = \frac{e^{-c} c^k}{k!}$ \cite{bollobas}.


\section{The replica approach} \label{sec3}

According to eq. (\ref{fra}), the characteristic function is obtained  
by calculating  the moments of the generating function. Substituting eq. (\ref{ewk}) in
eq. (\ref{fra}), we can rewrite
\begin{align}
&\mathcal{G}_N (\mu,\lambda) = \lim_{\epsilon \rightarrow 0^{+}}  \lim_{n_{\pm} \rightarrow \pm \frac{\mu}{\pi} }    
\left( \frac{-i}{2 \pi} \right)^{\frac{N n_{-}}{2}} \left( \frac{i}{2 \pi} \right)^{\frac{N n_{+}}{2}} 
\mathcal{D}_{n_{\pm} }(z),
\label{72sd}
\end{align}
in which we have defined the function
\begin{equation}
\mathcal{D}_{n_{\pm} }(z) = 
\int \left( \prod_{i=1}^{N}  d\bphi_i d \bpsi_i   H_{z}(\bphi_i,\bpsi_i )  \right)  
\mathcal{F}(\{ \bphi_{i}, \bpsi_{i}  \}), \label{dewa}
\end{equation}
with 
\begin{align}
&H_{z}(\bphi, \bpsi) = \exp{\left( - \frac{i z}{2} \bphi^2 +  \frac{i z^{*}}{2} \bpsi^2   \right)},  \nonumber \\
&\mathcal{F}(\{ \bphi_{i}, \bpsi_{i}  \}) = \left\langle  \exp{\left(  i \sum_{i < j} c_{ij} J_{ij} \left(  \bphi_{i}.\bphi_{j} - \bpsi_{i}.\bpsi_{j}   \right)     \right)  }    \right\rangle \nonumber .
\end{align}
The objects $\bphi_i = (\phi_{i}^{1},\dots,\phi_{i}^{n_{-}})$ and $\bpsi_i = (\psi_{i}^{1},\dots,\psi_{i}^{n_{+}})$ are the replicated
vectors at node $i$. The ensemble average $ \left\langle \dots  \right\rangle$  includes the average over the 
distribution of $\{ c_{ij} \}$, defined in eq. (\ref{prs}), and the average over the 
weights $\{ J_{ij} \}$, whose distribution $P(J)$ is arbitrary. In this section
we evaluate the leading term of $N^{-1} \ln \mathcal{D}_{n_{\pm} }(z)$  for $N \rightarrow \infty$ by means of the saddle-point method.

Using an integral
representation for the Kronecker delta in eq. (\ref{prs}), the average
over the topological disorder is explicitly calculated and the function $\mathcal{F}$
 reads
\begin{align}
&\mathcal{F}(\{ \bphi_{i}, \bpsi_{i}  \}) = \frac{e^{- \frac{N c}{2} }  }{C_{N}} \int_{0}^{2 \pi} \left( \prod_{i=1}^{N} \frac{d x_i}{2 \pi}  e^{i k_i x_i}  \right) \nonumber \\ 
&\times
\exp{\left( \frac{c}{2 N} \sum_{ij=1}^{N} e^{-i (x_i + x_j )} A(\bphi_{i}, \bpsi_{i};\bphi_{j}, \bpsi_{j} )      \right)}\,,  
\label{kjds}
\end{align}
where
\begin{equation}
A(\bphi, \bpsi;\bphi^{\prime},\bpsi^{\prime}) =   \left\langle \exp{\left[ i J \left( \bphi.\bphi^{\prime} -  \bpsi.\bpsi^{\prime}   \right) \right]} \right\rangle_{J}  \,,
\end{equation}
and $\langle \dots  \rangle_J$ stands for the average over $J$. We have 
retained only the leading contribution of $O(N)$ in the exponent of eq. (\ref{kjds}). 
To proceed further, the order-parameter 
\begin{equation}
\rho(\bphi, \bpsi) = \frac{1}{N} \sum_{i=1}^{N} e^{-i x_i} \delta(\bphi - \bphi_i ) \delta(\bpsi - \bpsi_i )
\end{equation}
is introduced in eq. (\ref{kjds}) by means of a functional delta, yielding the expression 
\begin{align} 
&\mathcal{F}(\{ \bphi_{i}, \bpsi_{i}  \}) = \frac{e^{- \frac{N c}{2} }  }{C_{N}} 
\int \mathcal{D}\rho \mathcal{D}\hat{\rho} \nonumber \\ 
&\times
\exp{\left( i N \int d \bphi \, d \bpsi \,  \rho(\bphi,\bpsi) \hat{\rho}(\bphi,\bpsi)    \right)} \nonumber \\ 
&\times
\exp{\left( \frac{c N}{2} \int d \bphi \, d \bpsi \, \rho(\bphi,\bpsi)   r(\bphi,\bpsi)     \right)} \nonumber \\
&\times
 \int \left( \prod_{i=1}^{N} \frac{d x_i}{2 \pi}    e^{i k_i x_i}  \right)  
\exp{\left( - i \sum_{i=1}^{N} e^{- i x_i} \hat{\rho}(\bphi_i,\bpsi_i) \right)} ,
\label{gaq}
\end{align}
with
\begin{equation}
r(\bphi, \bpsi) = \int d \bphi^{\prime} d \bpsi^{\prime}A(\bphi, \bpsi;\bphi^{\prime}, \bpsi^{\prime} ) \rho(\bphi^{\prime},\bpsi^{\prime}).
\end{equation}
The conjugated order parameter $\hat{\rho}(\bphi,\bpsi)$ has been rescaled according
to $\hat{\rho}(\bphi,\bpsi) \rightarrow N \hat{\rho}(\bphi,\bpsi)$ and the functional measure in the above integral
may be written as $\mathcal{D}\rho \mathcal{D}\hat{\rho} = \prod_{\bphi,\bpsi} \frac{N}{2 \pi} d \rho(\bphi,\bpsi) d \hat{\rho}(\bphi,\bpsi)$, where
the product runs over all possible values of $\bphi$ and $\bpsi$.
By substituting the large-$N$ leading contribution to $C_N$ in eq. (\ref{gaq})  
\begin{equation}
 C_N = \exp{\left[ N   \left( c \ln c - c - \sum_{k=0}^{\infty} p_k  \ln{k!}    \right) + O(1) \right] }\, ,
\end{equation}
and then inserting the resulting expression into
eq. (\ref{dewa}), we arrive at the integral form
\begin{equation}
\mathcal{D}_{n_{\pm} }(z)   =
 \int \mathcal{D}\rho \mathcal{D}\hat{\rho} 
\exp{\left( N S[\rho,\hat{\rho}]   \right)  } \,,
\label{suw}
\end{equation}
where the action reads
\begin{align}
&S[\rho,\hat{\rho}] = \frac{c}{2} - c \ln c + i  \int  d \bphi \, d \bpsi \, \hat{\rho}(\bphi,\bpsi)  \rho(\bphi,\bpsi) \nonumber \\
 &+ \frac{c}{2} \int  d \bphi \, d \bpsi \,
\rho(\bphi,\bpsi)  r(\bphi,\bpsi) \nonumber \\
&+
\sum_{k=0}^{\infty} p_k \ln \Bigg{\{} \int  d \bphi \, d \bpsi \, H_{z}(\bphi, \bpsi) 
\left[ - i  \hat{\rho}(\bphi,\bpsi)  \right]^{k} \Bigg{\}} . \label{wqk}
\end{align}

The integral in eq. (\ref{suw}) can be suitably evaluated through
the saddle-point method. In the limit $N \rightarrow \infty$, the function
$\mathcal{D}_{n_{\pm} }(z) $ is given by
\begin{equation}
\mathcal{D}_{n_{\pm} }(z)   \sim \exp{\left( N S[\rho,\hat{\rho}]   \right)  },
\label{jqa}
\end{equation}
where the order-parameters $\rho(\bphi,\bpsi)$ and $\hat{\rho}(\bphi,\bpsi)$ 
fulfill the saddle-point equations
\begin{align}
&\hat{\rho}(\bphi,\bpsi)  = i \, c \, r(\bphi, \bpsi) , \label{wwdw} \\
&\rho(\bphi,\bpsi) = \sum_{k=0}^{\infty} \frac{k p_k }{c} \frac{H_{z}(\bphi, \bpsi) 
\left[  r(\bphi, \bpsi)    \right]^{k-1} }
{\int d \bphi^{\prime}  d \bpsi^{\prime}  H_{z}(\bphi^{\prime}, \bpsi^{\prime})  \left[ r(\bphi^{\prime}, \bpsi^{\prime})   \right]^{k}   } .
\label{w1wdw}
\end{align}
Equations (\ref{wwdw}) and (\ref{w1wdw}) are obtained by extremizing the action $S[\rho,\hat{\rho}]$
with respect to $\rho$ and $\hat{\rho}$, respectively.
Inserting eqs. (\ref{wwdw}) and (\ref{w1wdw}) back into eq. (\ref{wqk}) and 
noting from eq. (\ref{w1wdw}) that
\begin{equation}
\int   d \bphi \, d \bpsi \, \rho(\bphi,\bpsi) r(\bphi, \bpsi) = 1 \nonumber ,
\end{equation}
we derive the compact expression 
\begin{equation}
S[\rho,\hat{\rho}] = \sum_{k=0}^{\infty} p_k \ln \Bigg{\{} \int  d \bphi \, d \bpsi \, H_{z}(\bphi, \bpsi) 
\left[ r(\bphi,\bpsi)  \right]^{k} \Bigg{\}} \,.
\label{ksi}
\end{equation}
The last step consists in performing the 
limit $n_{\pm} \rightarrow  \pm \frac{\mu}{\pi}$ in the above equation.
In order to make progress in this task, we need to make an assumption regarding the structure of
$\rho(\bphi,\bpsi)$ in the replica space. 


\section{The characteristic function of the index distribution} \label{sec4}

We follow previous works \cite{Dean,K08} and, with a modest amount
of foresight, we assume that $\rho(\bphi,\bpsi)$ has the following 
Gaussian form
\begin{align} 
&\rho(\bphi,\bpsi) = \frac{1}{U(n_{\pm})} \int d u \, d v \, W_{n_{\pm}}(u,v) \left(  \frac{i}{2 \pi u}   \right)^{\frac{n_{-}}{2}} \nonumber \\
&\qquad \qquad \times
\left(  \frac{i}{2 \pi v}   \right)^{\frac{n_{+}}{2}}  
\exp{\left( - \frac{i}{2 u} \bphi^2  - \frac{i}{2 v} \bpsi^2     \right)},
\label{sdrf} 
\end{align}
where $W_{n_{\pm}}(u,v)$ is the normalized joint distribution of the
complex variances $u$ and $v$, with ${\rm Im} \, u > 0$
and ${\rm Im} \, v > 0$. The latter conditions ensure the convergence of the integrals 
in eq. (\ref{sdrf}).
Since $\rho(\bphi,\bpsi)$ is not normalized
for arbitrary $n_{\pm}$ (see  eq. (\ref{w1wdw})), the 
factor $U(n_{\pm})$ has been consistently included in eq. (\ref{sdrf}).
The above replica symmetric (RS) form of $\rho(\bphi,\bpsi)$ 
remains invariant under rotations of the vectors $\bphi$ and $\bpsi$ as well
as under permutations of the vector components. 
A rigorous approach \cite{bordenave} for the eigenvalue distribution of sparse random graphs 
has confirmed the exactness of the results
obtained via the RS assumption.

By inserting eq. (\ref{sdrf}) in eq. (\ref{w1wdw}) and then taking
the limit $n_{\pm} \rightarrow  \pm \frac{\mu}{\pi}$, one derives the following equations for  
$W_{\mu}(u,v)$ and  $U(\mu)$
\begin{align}
&W_{\mu}(u,v) = \left[ U(\mu) \right]^2 \sum_{k=0}^{\infty} \frac{k p_k}{c} \frac{Q_{\mu}(u,v|k-1) \left( v/u   \right)^{\frac{\mu}{2 \pi}}  }
{ \int d u\, dv \, Q_{\mu}(u,v|k) \left( v/u   \right)^{\frac{\mu}{2 \pi}}  },  \label{gfes} \\
&\left[ U(\mu) \right]^{-2}  = \sum_{k=0}^{\infty} \frac{k p_k}{c} \frac{\int d u \, d v \, Q_{\mu}(u,v|k-1) \left( v/u   \right)^{\frac{\mu}{2 \pi}}  }
{ \int d u \, dv \, Q_{\mu}(u,v|k) \left( v/u   \right)^{\frac{\mu}{2 \pi}}  },  \nonumber
\end{align}
where
\begin{align}
&Q_{\mu}(u,v|k) = \int \left( \prod_{r=1}^{k}d u_r \,   d v_r\,  d J_r\, W_{\mu}(u_r,v_r) \, P(J_r)  \right) \nonumber \\
&\times 
\delta\left[ u - \frac{ 1 }{\left( z - \sum_{r=1}^{k} J_{r}^{2} u_r \right) }   \right] 
   \delta\left[ v + \frac{ 1 }{\left(z^{*} + \sum_{r=1}^{k} J_{r}^{2} v_r \right) }   \right]  
\end{align}
is the conditional distribution of $u$ and $v$ for a given degree $k$. 
Finally, we substitute eq. (\ref{sdrf}) in eq. (\ref{ksi}) and perform
the limit $n_{\pm} \rightarrow  \pm \frac{\mu}{\pi}$, from which the expression for the
large $N$ behavior of $\mathcal{G}_N (\mu,\lambda)$ is derived
\begin{align}
&\mathcal{G}_N (\mu,\lambda) = \lim_{\epsilon \rightarrow 0^{+}}  
\exp{\Big{\{} - \frac{N c}{2} \ln \left[U(\mu)\right]^{2}   \Big{\}} } \nonumber \\
&\times
\exp{\Bigg{\{} N \sum_{k=0}^{\infty} p_k \ln{\left[ \int d u \, d v \, Q_{\mu} (u,v|k) \left(- \frac{v}{u}   \right)^{\frac{\mu}{2 \pi}}  \right]  }     \Bigg{\}} } \,.
\label{pqfr}
\end{align}
In principle, eq. (\ref{pqfr}) determines completely the large-$N$ behavior of the characteristic function
for the index distribution 
of random graphs with arbitrary degree and edge distributions,
as long as a solution for $W_{\mu}(u,v)$
is extracted from the intricate self-consistent equation (\ref{gfes}).

For $\lambda=0$, one can show that $W_{\mu} (u,v) = \delta(u-v) R_{\mu} (u)$ solves
eq. (\ref{gfes}), provided the normalized distribution $R_{\mu} (u)$ fulfills
a certain equation, whose particular form is not relevant in this case.
Thus, the characteristic function at $\lambda=0$ simply reads
\begin{equation}
\mathcal{G}_N (\mu,0) = \exp{\left( \frac{i \mu N}{2} \right)},
\label{jjss2}
\end{equation}
which yields the delta peak  $\mathcal{P}_N (K,0) = \delta\left[ K - N/2 \right]$ for the index 
distribution, after substituting eq. (\ref{jjss2}) in eq. (\ref{fg1}).
This result reveals that, in order to access
the index fluctuations in this case, one needs to compute the 
next-order contribution to $\mathcal{G}_N (\mu,0)$ for large $N$.
The same situation arises in the replica approach for the GOE ensemble \cite{Cavagna}.
We present in the next section explicit results for the mean and the variance of
the index for specific  random graph models in the regime $|\lambda| > 0$. 


\section{Statistical properties of the index} \label{sec5}

It is straightforward
to check from eqs. (\ref{mom}) and (\ref{pqfr}) that the moments $\langle K^{n} \rangle$
scale as $\langle K^{n} \rangle \propto N^n$ for large $N$. In particular, the mean
and the variance read
\begin{align}
\langle K \rangle = N m(\lambda), \nonumber \\
\langle K^{2} \rangle -  \langle K \rangle^{2} = N \sigma^2(\lambda), \label{jqcv}
\end{align}
where the prefactors $m(\lambda)$ and $\sigma^2(\lambda)$ 
depend on the specific graph ensemble via
the distributions $p_k$ and $P(J)$.
Equation (\ref{jqcv}) differs strikingly
from rotationally invariant
ensembles of random matrices \cite{Cavagna,Majumdar1,Majumdar2,Vivo,Majumdar3}, where the variance of the typical index
fluctuations is of $O(\ln N)$ and the prefactor is independent of $\lambda$ \cite{Cavagna,Majumdar1,Majumdar2,Vivo}.
From eq. (\ref{jjss2}) we conclude that $\sigma^{2}(0) = 0$, which suggests that the index variance
of random graphs with an arbitrary degree distribution
exhibits the logarithmic scaling $\langle K^{2} \rangle -  \langle K \rangle^{2} \propto \ln N$
for large $N$ at this particular $\lambda$. This is confirmed below for the case of ER random graphs
by means of numerical diagonalization results.

For $|\lambda| > 0$, the intensive quantities $m(\lambda)$ and $\sigma^{2}(\lambda)$ are obtained directly
from eqs. (\ref{mom}) and (\ref{pqfr}), i.e., from the coefficients of the expansion of $\mathcal{G}_N (\mu,\lambda)$ around $\mu=0$. 
In general, $m(\lambda)$ and $\sigma^{2}(\lambda)$ are given in terms
of averages with the distribution
$W_{0}(u,v) = \lim_{\mu \rightarrow 0} W_{\mu}(u,v)$, whose self-consistent
equation is derived by performing the limit $\mu \rightarrow 0$ in eq. (\ref{gfes})
\begin{equation} 
W_{0}(u,v) =  \sum_{k=0}^{\infty} \frac{k p_k}{c} Q_{0}(u,v|k-1) . 
\label{hspc}
\end{equation}
The object $W_{0}(u,v)$ may be interpreted as the averaged joint distribution of the 
diagonal resolvent elements at the two different points $z$ and $-z^{*}$ of the complex plane.
The resolvent
elements at $z$ and $-z^{*}$ are both calculated on the same cavity graph \cite{Metz2010,Biroli}, defined as the graph in which an 
arbitrary node and all its edges are deleted. 

Equation (\ref{hspc}) has a simpler form when compared to eq. (\ref{gfes}) and numerical solutions
for $W_{0}(u,v)$ can be obtained using the population dynamics algorithm \cite{K08}, where the distribution $W_{0}(u,v)$ is parametrized by a large
set $\{ u_i,v_i \}_{i=1,\dots,M}$ containing $M$ pairs of stochastic random variables.
These are updated iteratively
according to their
joint distribution $W_{0}(u,v)$, governed by eq. (\ref{hspc}), until $W_{0}(u,v)$ attains a stationary profile.
The limit $\epsilon \rightarrow 0^{+}$ in eq. (\ref{pqfr}) is handled numerically by calculating
$W_{0}(u,v)$ for small but finite values of $\epsilon$.
We refer the reader to references \cite{K08,TimTese,Metz2010} for further details
regarding the population dynamics algorithm in the context of random matrices 
and some technical points involved in the limit $\epsilon \rightarrow 0^{+}$.
Since the eigenvalue distribution $\rho_N(\lambda)$ is symmetric around $\lambda=0$, 
$m(\lambda)$ and $\sigma^2(\lambda)$
obey the relations $m(-\lambda) = 1 - m(\lambda)$ and $\sigma^2(\lambda) =  
\sigma^2(-\lambda)$. Hence the results for $m(\lambda)$ and $\sigma^2(\lambda)$ discussed
below are limited to the sector $\lambda \geq 0$.


\subsection{Erd\"os-R\'enyi random graphs}

For ER random graphs the quantities $m(\lambda)$ and $\sigma^2(\lambda)$
read
\begin{align}
&m(\lambda) = \lim_{\epsilon \rightarrow 0^{+}} \Bigg{[}  \int d u \, d v \, d u^{\prime} \, d v^{\prime} \, W_{0}(u,v)  \nonumber \\
& \qquad \qquad \qquad \quad \times W_{0}(u^{\prime},v^{\prime}) \Delta_1 (u,v;u^{\prime},v^{\prime} ) \Bigg{]},\label{msa}
\end{align}
\begin{align}
&\sigma^{2}(\lambda) = \lim_{\epsilon \rightarrow 0^{+}} \Bigg{[}  \int d u \, d v \, d u^{\prime} \, d v^{\prime} \, W_{0}(u,v)  \nonumber \\
& \qquad \qquad \qquad \quad \times W_{0}(u^{\prime},v^{\prime}) \Delta_2 (u,v;u^{\prime},v^{\prime} ) \Bigg{]},  \label{msa1}
\end{align}
where
\begin{align}
&\Delta_1 (u,v;u^{\prime},v^{\prime} ) = \frac{i c}{4 \pi} \left\langle  F_J (u,v;u^{\prime},v^{\prime})      \right\rangle_{J} 
- \frac{i}{2 \pi}  \ln{\left( - \frac{v}{u}   \right)},  \nonumber \\
&\Delta_2 (u,v;u^{\prime},v^{\prime} ) = \frac{c}{8 \pi^{2}} \left\langle \left[  F_J (u,v;u^{\prime},v^{\prime})   \right]^{2}   \right\rangle_{J} \nonumber \\
& + \frac{1}{4 \pi^{2} }  \ln{\left( - \frac{v}{u}   \right)} \ln{\left( - \frac{v^{\prime}}{u^{\prime}}   \right)} - \frac{1}{4 \pi^{2} }  \left[ \ln{\left( - \frac{v}{u}   \right)} \right]^{2}, \nonumber
\end{align}
with
\begin{equation}
F_J (u,v;u^{\prime},v^{\prime}) = \ln{\left(\frac{1 - J^2 u u^{\prime}  }{1 - J^2 v v^{\prime}} \right)}.
\end{equation}
The distribution $W_{0}(u,v)$ is calculated numerically from eq. (\ref{hspc}) using the population dynamics
algorithm with the degree
distribution $p_k = \frac{e^{-c} c^{k}}{k!}$ of ER random graphs \cite{bollobas}. 
 
\begin{figure}[t!]
\center
\includegraphics[scale=0.9]{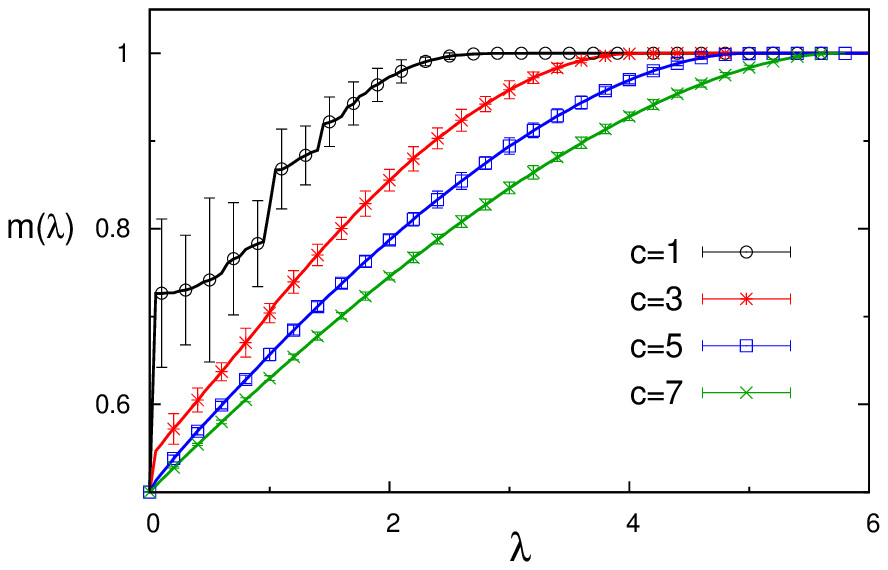}
\caption{
Numerical results for the averaged intensive index $m(\lambda)$ of Erd\"os-R\'enyi random graphs
with the distribution of edges $P(J) = \delta(J-1)$, obtained
using the population dynamics algorithm (solid lines) with $M = 10^{6}$ random variables
and $\epsilon= 10^{-3}$. Numerical diagonalization results (symbols), calculated from an
ensemble of $100$ matrices of size $N = 3200$, are shown as a comparison.
 }
\label{compma}
\end{figure}

\begin{figure}[t!]
\center
\includegraphics[scale=0.9]{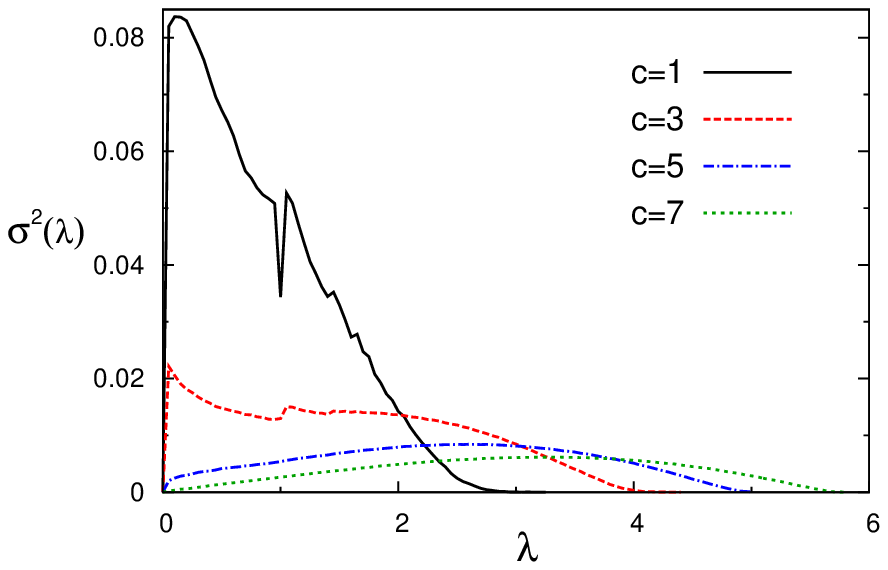}
\caption{
Numerical results for the prefactor $\sigma^{2}(\lambda)$ of the index variance of  Erd\"os-R\'enyi random graphs with the distribution of edges $P(J) = \delta(J-1)$, obtained
using the population dynamics algorithm with $M = 10^{6}$ random variables
and $\epsilon= 10^{-3}$.
 }
\label{compDiag}
\end{figure}

In figures \ref{compma} and \ref{compDiag}, we present numerical results for $m(\lambda)$ and $\sigma^2(\lambda)$
in the case of ER random graphs with $P(J) = \delta(J - 1)$. The discontinuous behavior of
$m(\lambda)$ for small average degree $c$ reflects the presence of delta peaks in the eigenvalue distribution, due to
the proximity of the percolation transition \cite{bauer}. In fact, all connected
components of ER random graphs are finite trees and the spectrum is purely discrete for $c < 1$, while
the heights of these peaks decrease exponentially with increasing $c$ \cite{bauer}.
The calculation of the integrated density of states presented here allows to determine, for $N \rightarrow \infty$, not only the location 
of the most important delta peaks 
in the spectrum, but also their relative weights, given by the size of the discontinuities of $m(\lambda)$.  The 
exactness of our results for  $m(\lambda)$ is confirmed by the comparison with numerical
diagonalization data, as shown in figure \ref{compma}.

\begin{figure}[t!]
\centering
\subfigure[Index variance for $\lambda > 0$. The solid lines represent the linear fit $\langle K^{2} \rangle -  \langle K \rangle^{2} = a + b N$, with the values
of the slope $b$ indicated next to each straight line. The theoretical values for $\sigma^{2}(\lambda)$, calculated
through the numerical solution of eq. (\ref{msa1}), are given by $\sigma^{2}(0.5) = 0.015$, $\sigma^{2}(3.0) = 0.0085$
and $\sigma^{2}(3.5) = 0.0040$.]{
\includegraphics[scale=0.9]{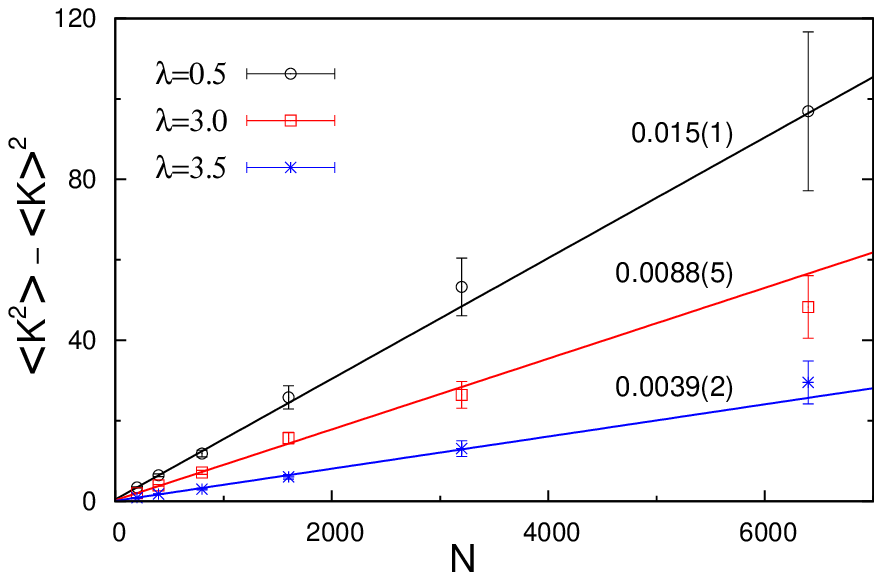}
\label{sub1}
}
\subfigure[ Index variance for $\lambda = 0$. The solid line represents the logarithmic fit $\langle K^{2} \rangle -  \langle K \rangle^{2} = a + b \ln N$, with the
slope $b = 0.47(6)$.]{
\includegraphics[scale=0.9]{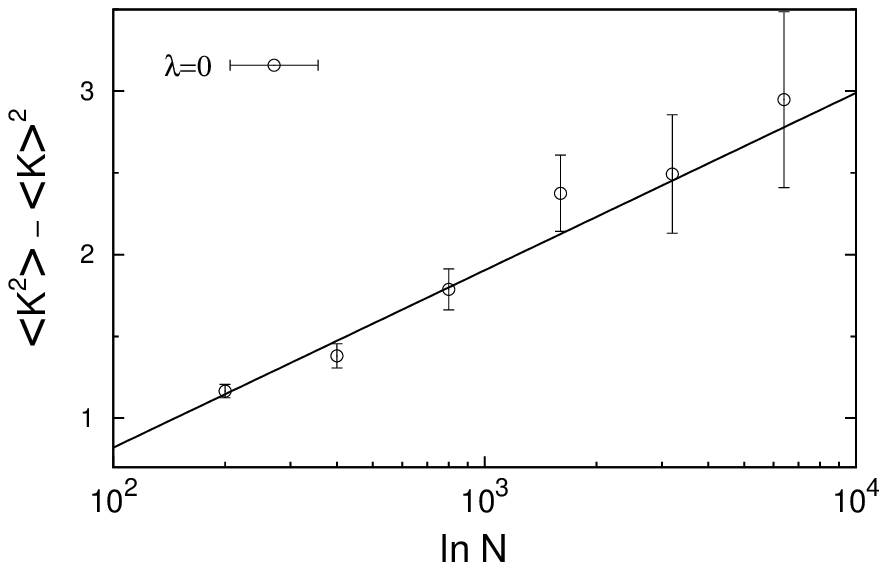}
\label{sub2}
}
\caption{Numerical diagonalization results for the index variance of Erd\"os-R\'enyi random graphs with $c=3$ as
a function of the number of nodes $N$. Each data point is calculated 
from an ensemble with $S$ independent realizations of the adjacency matrix $\bA$, where
$S$ has been chosen according to $S = \frac{3.2 \times 10^{5}}{N}$. The solid lines represent
the best fits of the numerical data. 
}
\label{fklk}
\end{figure}

The results for the prefactor $\sigma^2(\lambda)$ of ER random graphs are shown in figure \ref{compDiag}. For the smaller
values of $c$, the index fluctuations are generally stronger and  $\sigma^2(\lambda)$ exhibits an irregular
behavior, both features related to strong sample to sample fluctuations of the graph structure close
to the percolation critical point. The prominent feature of figure \ref{compDiag} is that $\sigma^2(\lambda)$ shows 
a non-monotonic behavior, with a maximum for a certain intermediate value
of $\lambda$ and a vanishing behavior at $\lambda = 0$, which signals
the breakdown of the linear scaling $\langle K^{2} \rangle -  \langle K \rangle^{2} \propto N$. 
This is confirmed by the numerical diagonalization results
of figure \ref{fklk}, where $\langle K^{2} \rangle -  \langle K \rangle^{2}$ is calculated
as a function of $N$ for $c=3$. 
The results of figure \ref{sub1}, for different values of $\lambda > 0$, display a linear behavior for increasing $N$, with slopes in full accordance
with the theoretical values for $\sigma^2(\lambda)$, as indicated on the caption. On the other hand, figure
\ref{sub2} shows that the index variance scales as $\langle K^{2} \rangle -  \langle K \rangle^{2} \propto \ln N$ for
$\lambda=0$, similarly to the behavior of rotationally invariant ensembles \cite{Cavagna,Majumdar1,Majumdar2,Vivo,Majumdar3}.


\subsection{Random regular graphs}

In the case of random regular graphs, the degree distribution
is simply $p_k = \delta_{k,c}$ \cite{wormald}, where $c > 2$ is an integer. Firstly, let us
consider the situation in which the values of the edges are
fixed, i.e., their distribution reads $P(J^{\prime}) = \delta(J^{\prime}-J)$, with $J \in \mathbb{R}$. In this
case, eq. (\ref{gfes}) has the following solution for arbitrary $\mu$
\begin{equation}
W_{\mu}(u,v) = \left( - \frac{g \, v  }{g^{*}  u  }    \right)^{\frac{\mu }{2 \pi }} 
\delta\left( u - g   \right) \delta\left( v +  g^{*}  \right),
\label{hqnn}
\end{equation}
where $g$ is a root of the algebraic equation
\begin{equation}
(c-1) J^2 g^2 - z g + 1 = 0 .
\label{ffvvb}
\end{equation}
The quantity $g$ represents the diagonal elements of the 
resolvent on the cavity graph \cite{Metz2010,Biroli}.
Substituting eq. (\ref{hqnn}) in eq. (\ref{pqfr}) and using
the above quadratic equation, we get
\begin{equation}
\mathcal{G}_N (\mu,\lambda) = \lim_{\epsilon \rightarrow 0^{+}}  
\exp{\left[ i \mu N m(z) \right]},
\label{jjss}
\end{equation}
where 
\begin{equation}
m(z) = \frac{1}{\pi} {\rm Im}\left[\ln{\left( z - c J^2 g   \right)} \right] - \frac{c}{2 \pi} {\rm Im}\left[\ln{\left( 1 - J^2 g^2   \right)}\right].
\end{equation}
Equation (\ref{jjss}) is the large-$N$ behavior of $\mathcal{G}_N (\mu,\lambda)$ for random regular
graphs in the absence of edge fluctuations. By choosing the proper roots of eq. (\ref{ffvvb})
in the different sectors of the spectrum \cite{Metzfinite}, we can perform
the limit $\lim_{\epsilon \rightarrow 0^{+}} m(z)$ and derive the following analytical result 
for $\lambda \geq 0$
\begin{align}
m(\lambda) &= 1 + \frac{1}{\pi} \tan^{-1}{\left[\frac{-c  \sqrt{\lambda_{b}^{2} - \lambda^{2}} }{\lambda (c-2) }    \right]  } \nonumber \\
&- \frac{c}{2 \pi} \tan^{-1}{\left[  \frac{\lambda \sqrt{\lambda_{b}^{2} - \lambda^{2}}  }{\lambda^{2} - 2 c (c-1) J^2 }     \right]  },
\label{hhgq}
\end{align}
with $|\lambda_b| = 2 |J| \sqrt{c-1}$ denoting the band edge of the 
continuous spectrum of random regular graphs \cite{Kesten,McKay}.
Equation (\ref{hhgq}) coincides with the average integrated density of states
in the bulk of a Cayley tree \cite{Derrida} and it converges to the result 
for the GOE ensemble when $c \gg 1$ \cite{Cavagna}, as long
as we rescale $J$ according to $J \rightarrow J/\sqrt{c}$.
The substitution of eq. (\ref{jjss}) in eq. (\ref{fg1}) yields
a delta peak  $\mathcal{P}_N (K,\lambda) = \delta\left[ K - N m(\lambda)  \right]$, which
implies that $\sigma^{2} (\lambda) = 0$. This suggests  that the index variance
exhibits the logarithmic scaling $\langle K^{2} \rangle -  \langle K \rangle^{2} \propto \ln N$
for arbitrary $\lambda$.
The latter property is consistent with the absence 
of localized states and the corresponding 
repulsion between nearest-eigenvalues, which is common
to the whole spectrum of random regular graphs with uniform edges 
\cite{jakobson, Smilansky,Geisinger}.
\begin{figure}[t!]
\centering
\subfigure[]{
\includegraphics[scale=0.9]{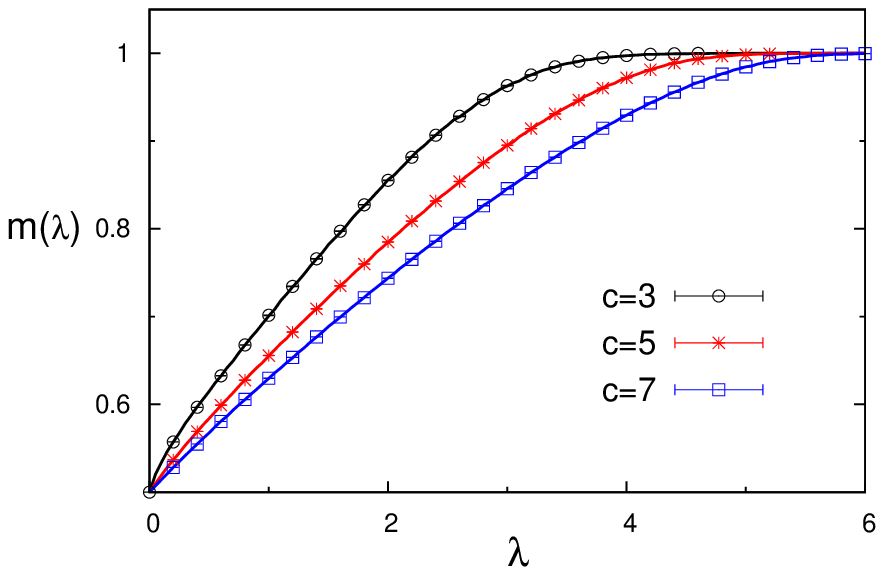}}
\subfigure[]{
\includegraphics[scale=0.9]{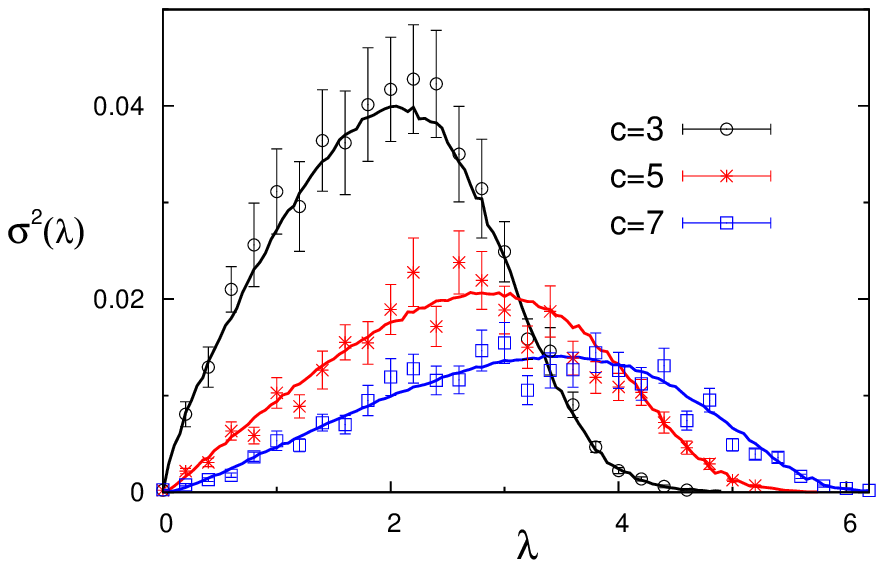}}
\caption{Numerical results for the averaged intensive index $m(\lambda)$ and
the prefactor $\sigma^{2}(\lambda)$ of the index variance of random regular graphs with edges
drawn from the Gaussian distribution $P(J) = \left( 2 \pi  \right)^{-\frac{1}{2}}
\exp{\left( -J^2/2 \right)} $, obtained
using the population dynamics algorithm (solid lines) with $M = 5 \times 10^{5}$ random variables
and $\epsilon= 10^{-3}$. Numerical diagonalization results (symbols), calculated from an
ensemble of $100$ matrices of size $N = 4000$, are shown as a comparison.
}
\label{fkwq}
\end{figure}

The above results are clearly due to our trivial choice for $P(J)$. The spectrum
of random regular graphs contains localized states in the 
presence of edge disorder \cite{K08,semerjian} and one can expect that
$\sigma^{2}(\lambda)$ exhibits a nontrivial behavior as long as $P(J)$ has a finite variance.
The functions $m(\lambda)$ and  $\sigma^{2}(\lambda)$ for random regular graphs
with an arbitrary distribution $P(J)$ read
\begin{align}
&m(\lambda) = \frac{i}{2 \pi} \lim_{\epsilon \rightarrow 0^{+}} \left[\frac{c}{2} K_{1}(z) - L_{1}(z)     \right], \nonumber \\
&\sigma^{2}(\lambda) = \frac{1}{4 \pi^2} \lim_{\epsilon \rightarrow 0^{+}} \Bigg{\{} \frac{c}{2}  K_{2}(z) -  \frac{c}{2}  \left[ K_{1}(z) \right]^{2} \nonumber \\
& \qquad \qquad \qquad \qquad + \left[ L_{1}(z) \right]^{2} - L_{2}(z)    \Bigg{\}},
\end{align}
where $K_{n}(z)$ and $L_{n}(z)$ are calculated from
\begin{align}
K_{n}(z) &= \int d u \, d v \, d u^{\prime} \, d v^{\prime}
Q_{0}(u,v|c-1) Q_{0}(u^{\prime},v^{\prime}|c-1)  \nonumber \\
&\times \left\langle  \left[ F_J (u,v;u^{\prime},v^{\prime}) \right]^{n}  \right\rangle_{J}, \nonumber \\
L_{n}(z)  &= \int d u \, d v \, Q_{0}(u,v|c) \left[ \ln{\left(- \frac{v}{u}  \right)}  \right]^{n}.
\end{align}
Figure \ref{fkwq} shows population dynamics results for $m(\lambda)$ and  $\sigma^{2}(\lambda)$
in the case of a Gaussian distribution $P(J) = \left( 2 \pi  \right)^{-\frac{1}{2}}
\exp{\left( -J^2/2 \right)} $.
The function $m(\lambda)$ does not display any 
noticeable discontinuity, as
observed previously for ER random graphs, 
due to the absence of disconnected clusters in the
case of large random regular graphs \cite{wormald}. In addition, we note that $\sigma^{2}(\lambda)$ has qualitatively the same
non-monotonic behavior as in ER random graphs, exhibiting a maximum for a certain $\lambda$ and 
approaching zero as $\lambda \rightarrow 0$. Numerical diagonalization 
results for large matrices $\bA$, also shown in figure \ref{fkwq}, confirm the correctness of our theoretical approach.


\subsection{The index distribution}

In this subsection, we inspect the full index distribution of random graphs using numerical
diagonalization, instead of undertaking the more difficult task of
calculating the characteristic function from the numerical solution
of eqs. (\ref{gfes}) and (\ref{pqfr}). We restrict ourselves to $\lambda > 0$, where the index variance scales linearly with $N \gg 1$. 
\begin{figure}[t!]
\centering
\subfigure[Erd\"os-R\'enyi random graphs with the distribution of the edges $P(J) = \delta(J -1)$.]{
\includegraphics[scale=0.9]{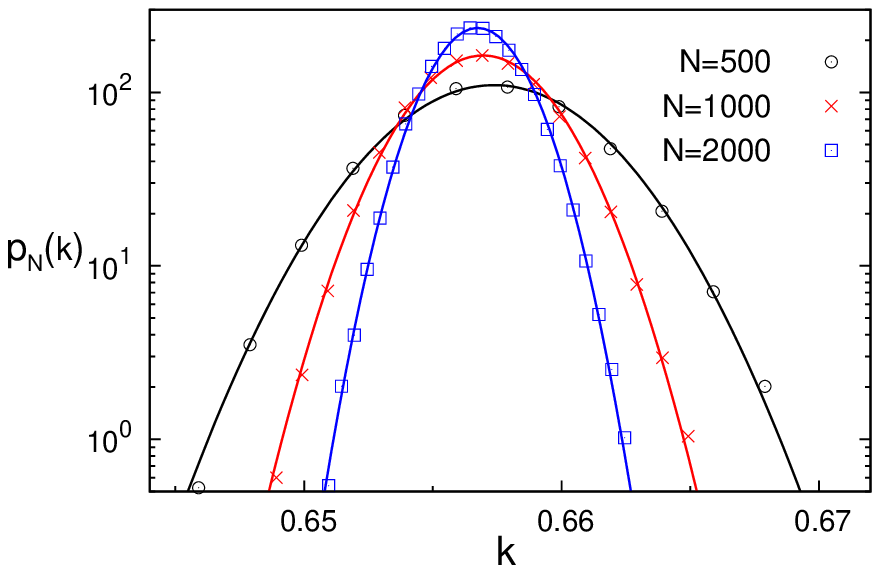}}
\subfigure[Regular random graphs with the distribution of the edges $P(J) = \left( 2 \pi  \right)^{-\frac{1}{2}} \exp{\left( -J^2/2 \right)}$.]{
\includegraphics[scale=0.9]{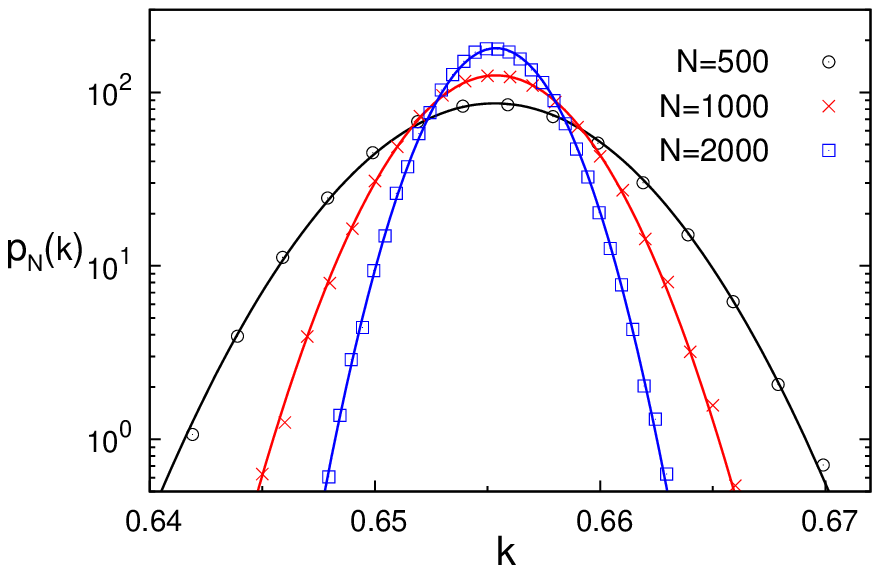}}
\caption{Numerical diagonalization results (symbols) for the distribution of the intensive index 
of random graphs 
with $c=5$ and $\lambda=1$. The histograms were generated
from  $10^{5}$ independent samples for the intensive index of the adjacency matrix $\bA$. The solid lines are Gaussian
distributions with mean and variance taken from the data.
}
\label{fklk1}
\end{figure}

In figure \ref{fklk1} we show results for the  distribution $p_N(k,\lambda)$
of the intensive index $k_N (\lambda)= \mathcal{K}_N(\lambda)/N$ in the case of ER and random regular graphs with $c=5$, obtained
from numerical diagonalization for $\lambda=1$. For each value of $N$, the results are compared with 
a Gaussian distribution (solid lines) with mean and variance taken from the data, which
confirms the Gaussian character of the typical index fluctuations for both random graph models 
when $N$ is large but finite. 

Overall, our results suggest
that, for $N \gg 1$ and $|\lambda| > 0$, the intensive index of ER and random regular graphs is distributed according
to
\begin{equation}
p_N(k,\lambda) = \sqrt{\frac{N}{2 \pi \sigma^{2}(\lambda)}} \exp{\Bigg{\{} -\frac{N}{2 \sigma^{2}(\lambda)} \left[ k - m(\lambda)  \right]^{2} \Bigg{\}} },
\label{jjw12}
\end{equation}
with non-universal parameters $\sigma^{2}(\lambda)$ and $m(\lambda)$ that depend
on the underlying random graph model as well as on the particular value of the threshold $\lambda$.
The function $p_N(k,\lambda)$ converges to 
$p_N(k,\lambda) = \delta{\left[ k - m(\lambda)   \right]}$
for $N \rightarrow \infty$, but the rate of convergence is slower  when compared to
rotationally invariant ensembles  \cite{Cavagna,Majumdar1,Majumdar2,Vivo,Majumdar3}, due to the logarithmic scaling of the 
index variance with respect to $N$ in the latter case. On the other hand, the Gaussian nature of the index fluctuations
for $N \gg 1$ seems to be an ubiquitous feature of random matrix models.


\section{Final remarks} \label{sec6}

We have presented an analytical expression for the characteristic
function of the index distribution describing a broad class of random graph models, which
comprises graphs with arbitrary degree and edge distributions. Ideally, this general result gives
access to all moments of the index distribution in the limit $N \rightarrow \infty$.
We have shown that the index variance of typical fluctuations is generally of $O(N)$, with a prefactor $\sigma^{2}(\lambda)$ that depends on the 
random graph model under study as well as on the threshold $\lambda$ that 
defines the index through eq. (\ref{i1}).
In particular, $\sigma^{2}(\lambda)$ follows an intriguing non-monotonic behavior for random graphs
with localized eigenstates: it exhibits a maximum at
a certain $|\lambda| > 0$ and a vanishing behavior at $\lambda= 0$.
Numerical diagonalization data confirm the theoretical results 
and support the Gaussian form of the typical index distribution for the random graphs
considered here (see eq. (\ref{jjw12})), completing
the picture about the index statistics.

Our results differ with those of rotationally invariant
ensembles, where the index variance is of $O(\ln N)$, with a prefactor that is
independent of $\lambda$ and has an universal character. We argue that
this difference in the scaling forms arises due to the presence of localized
states in the spectrum of some random graphs. In the localized sectors, the eigenvalues
do not repel each other and 
behave as uncorrelated random variables, such that
the total number of eigenvalues contained in finite regions within the localized phase 
suffers from stronger finite size fluctuations as compared to regions within the extended phase, 
where level-repulsion tends to equalize the space
between neighboring eigenvalues. On the other hand, the Gaussian
nature of typical index fluctuations seems to be a robust feature 
of random matrix models. 

On the methodological side, the replica approach as devised here departs from
the representation of the characteristic function in terms
of real Gaussian integrals, instead of the fermionic Gaussian integrals adopted in
reference \cite{Cavagna}. In the situations where $\sigma^{2}(\lambda) =0$, the logarithmic scaling of the index variance is obtained in our
setting from the next-to-leading order terms, for large $N$, in the saddle-point
integral of eq. (\ref{suw}). These contributions come from $O(1/\sqrt{N})$ fluctuations of the
order-parameter and they are handled following the ideas of reference \cite{Metzfinite}.
Indeed, we have precisely recovered
the analytical results for the GOE ensemble \cite{Cavagna} employing this strategy \cite{MetzGauss}, and 
the same approach can be used to calculate the prefactors in situations
where the variance of random graphs is of $O(\ln N)$.

Our work opens several perspectives in the study of the typical index
fluctuations. Firstly, it would be worth having  approximate 
schemes or numerical methods to solve eq. (\ref{gfes}) and obtain the distribution
$W_{\mu}(u,v)$, which would allow to fully determine the characteristic function for random graphs.
Due to the versatile character of the replica method, the study of
the averaged integrated density of states of the Anderson 
model on regular graphs \cite{Abouchacra73}  and its sample to sample fluctuations is just around 
the corner. 
It would be also interesting to inspect the robustness of the Gaussian form of the index fluctuations
in random matrix ensembles with strong inherent fluctuations, such
as Levy random matrices \cite{Bouchaud} and scale-free random networks \cite{barabasi}.
The index statistics of both random matrix models can be treated using the replica approach as developed here.
In fact, scale-free random graphs, crucial in modelling many real-world networks
appearing in nature \cite{Barrat}, can be studied directly from our work by choosing the degree distribution  as
$p_k \sim k^{-\gamma}$ ($2 < \gamma \leq 3$), which yields random graphs with 
strong sample to sample degree fluctuations. Finally, we point out that the different 
scaling behaviors of the index variance 
should have important consequences to 
the relaxation properties and search algorithms on complex energy surfaces.


\acknowledgements

FLM acknowledges the financial support from the Brazilian agency CAPES through the program Science
Without Borders.


\bibliography{bibliography.bib}

\end{document}